\documentclass[a4paper,12pt]{article}
\usepackage{wrapfig}
\usepackage{times}
\usepackage{geometry}             
\usepackage{graphicx}
\usepackage{amssymb}
\usepackage{epstopdf}
\def\gs{\mathrel{\raise0.35ex\hbox{$\scriptstyle >$}\kern-0.6em
\lower0.40ex\hbox{{$\scriptstyle \sim$}}}}
\def\ls{\mathrel{\raise0.35ex\hbox{$\scriptstyle <$}\kern-0.6em
\lower0.40ex\hbox{{$\scriptstyle \sim$}}}}

\newenvironment{bul2}
{\begin{list}
  {$\quad \bullet$}
  {\itemsep = 0.5ex\parsep=0pt\topsep = 1mm\leftmargin=8mm}}
{\end{list} }

\newcounter{numcnt}

\begin{document}

\centerline{\Large \bf HiZELS: the High Redshift Emission Line Survey with UKIRT}

\bigskip
\begin{center}
{\large{\it Philip Best$^1$, Ian Smail$^2$, David Sobral$^1$, Jim
    Geach$^2$, Timothy Garn$^1$, 
Rob Ivison$^3$ \\ 
Jaron Kurk$^4$, Gavin Dalton$^{5,6}$, Michele Cirasuolo$^{1,2}$, Mark
Casali$^7$\\}}
\bigskip

{\footnotesize\it
\begin{tabular}{ll}
1: & SUPA, Institute for Astronomy, Royal Observatory,
Blackford Hill, Edinburgh, EH9 3HJ, UK\\ 
2: & Institute of Computational Cosmology, Durham University, South Road,
Durham, DH1 3LE, UK\\ 
3: & Astronomy Technology Centre, Royal Observatory, Blackford
Hill, Edinburgh, EH9 3HJ, UK\\ 
4: & Max-Planck-Institut f\"ur Astronomie, K\"onigstuhl, 17 D-69117, Heidelberg, Germany\\
5: & Astrophysics, Department of Physics, Keble Road, Oxford, OX1 3RH, UK\\
6: & Space Science and Technology, Rutherford Appleton Laboratory, HSIC,
Didcot, OX11 0QX, UK\\
7: & European Southern Observatory, Karl-Schwarzschild-Strasse 2, D-85738 Garching, Germany\\
\end{tabular}
}  
\end{center}
\bigskip

\begin{center}
\parbox{15cm}{ {\large{\bf Abstract: }} In these proceedings we report on
HiZELS, the High-z Emission Line Survey, our successful panoramic
narrow-band Campaign Survey using WFCAM on UKIRT to detect and study
emission line galaxies at $z \simeq 1$--9.  HiZELS employs the H$_2$(S1)
narrow-band filter together with custom-made narrow-band filters in the
$J$ and $H$-bands, with the primary aim of delivering large
identically-selected samples of H$\alpha$ emitting galaxies at redshifts
of 0.84, 1.47 and 2.23. Comparisons between the luminosity function, the
host galaxy properties, the clustering, and the variation with environment
of these H$\alpha$-selected samples are yielding unique constraints on the
nature and evolution of star-forming galaxies, across the peak epoch of
star-formation activity in the Universe. We provide a summary of the
project status, and detail the main scientific results obtained so far:
the measurement of the evolution of the cosmic star-formation rate density
out to $z > 2$ using a single star-formation indicator, determination of
the morphologies, environments and dust-content of the star-forming
galaxies, and a detailed investigation of the evolution of their
clustering properties. We also summarise the on-going work and future
goals of the project.}
\end{center}
\bigskip
\smallskip

\noindent{\large{\bf 1. Introduction}}
\medskip

The fundamental observables required to understand the basic features of
galaxy formation and evolution are the volume-averaged star-formation rate
as a function of epoch, its distribution function within the galaxy
population, and the variation with environment.  Surveys of the
star-formation rate as a function of epoch suggest that the star-formation
rate density rises as $\sim (1+z)^{4}$ out to at least $z\sim1$ (e.g.\
Lilly et al.\ 1996), and then flattens, with the bulk of stars seen in
galaxies today having been formed between $z\sim 1$--3. Determining the
precise redshift where the star-formation rate peaked is more difficult,
however, with different star-formation indicators giving widely different
measures of the integrated star-formation rate density (see Hopkins et
al.\ 2006). These problems are exacerbated by the effects of cosmic
variance in the current samples, which are typically based on small-field
surveys.

The H$\alpha$ emission line is a very well-calibrated measure of
star-formation rate in the nearby Universe (e.g.\ Kennicutt 1998;
Moustakas et al.\ 2006). As it redshifts through the optical and near-IR
bands, it offers a single star-formation indicator which can be studied
from $z=0$ to $z\sim 3$, right through the peak star-formation epoch in
the Universe. It is relatively immune to dust extinction, and has
sufficient sensitivity that estimates of the integrated star-formation
rate don't require large extrapolations for faint sources below the
sensitivity limit: surveys with a sensitivity of
$\sim$\,10\,M$_\odot$\,yr$^{-1}$ can be undertaken in H$\alpha$ at
z\,$\sim$\,2 with current instrumentation, compared to limits of $\sim
10^2$--$10^3$\,M$_\odot$\,yr$^{-1}$ for other dust-independent tracers
such as radio, far-infrared and sub-mm luminosities (e.g.\ Ivison et al.\
2007).

The H$\alpha$ emission line has been widely used as a method of tracing
the evolution of star-formation, both through spectroscopic surveys and
via imaging surveys exploiting narrow-band filters (e.g.\ Gallego et al.\
1995; Yan et al.\ 1999; Tresse et al.\ 2002; Doherty et al.\
2006). Narrow-band surveys offer a sensitive and unbiased method of
detecting emission-line objects lying in large well-defined volumes; the
sources are identified on the strength of their emission line and thus
crudely represent a star-formation rate-selected sample, and they must lie
in a narrow range in redshift. Before the advent of large-area near-IR
detectors such as WFCAM, narrow-band H$\alpha$ surveys in the near-IR
(i.e.\ at $z \gs 0.7$) were limited to very small areas and sample sizes
(the largest at $z \sim 2$ had just $\sim 10$ candidate sources; Moorwood
et al.\ 2000). The primary goal of HiZELS is to overcome this, with
wide-area surveys using narrow-band filters in the $J$, $H$ and $K$-bands
to detect of order a thousand star-forming galaxies in H$\alpha$ at each
of three redshifts: 0.84, 1.47 and 2.23. These large samples, selected
with a uniform selection function, can be used to determine the H$\alpha$
luminosity function (LF) at each epoch, investigate any strong changes in
its shape, and provide the first reliable estimate of the change in the
global star-formation rate of the Universe between $z=0$ and 2.2 using a
single tracer of star-formation.

It is not only the global average star-formation rate which is important
for our understanding of galaxy formation and evolution, but more
crucially the nature and distribution of the star-forming galaxies at high
redshifts. Galaxies form and evolve within the hierarchically growing
dark-matter haloes of a $\Lambda$-CDM Universe, but the details of the
galaxy formation process depend upon the complicated gas dynamics of star
formation and feedback, and these processes are poorly understood. A
surprising result of many recent studies is that the stellar populations
of the most massive galaxies formed earlier than those of less massive
galaxies -- a process often referred to as ``downsizing'' (e.g.\ Cowie et
al.\ 1996). Massive galaxies must therefore form stars rapidly at an early
epoch, and then have their star-formation truncated, for example by
feedback from AGN (e.g.\ Bower et al.\ 2006; Best et al.\ 2006) -- but the
epoch at which this occurred is still uncertain. In the local Universe,
star-formation is also suppressed in dense environments (e.g.\ Lewis et
al.\ 2002, Best 2004); this effect diminishes with increasing redshift,
with hints that it disappears altogether at $z\sim 2$ (Kodama et al.\
2007). But where precisely, in terms of epoch and environment, does this
environmental influence begin to become important, and to what extent is
the build-up of galaxies into groups and clusters since $z \sim 1$
responsible for the sharp decline of the cosmic star-formation rate
density since $z \sim 1$?

HiZELS aims to tackle all of these issues by obtaining large samples of
star-forming galaxies in representative volumes at three epochs across the
peak epoch of star-formation in the Universe.  Coupled with lower redshift
studies, we are determining how the stellar mass of H$\alpha$-selected
galaxies declines with redshift between $z=2.2$ and $z=0$, investigating
the physical processes involved in the downsizing activity. We are
investigating changes in the H$\alpha$ luminosity function as a function
of environment at each epoch and between different epochs: sky areas of a
few square degrees are required to probe the full range of galaxy
environments at these redshifts. The samples are also large enough to give
a robust measurement of the clustering properties of sub-populations of
the H$\alpha$ emitters, split by other properties like mass or
star-formation rate; this provides important insights into their
properties, including information about the relative masses of their dark
matter haloes. Combining all of this information, HiZELS will provide a
strong test of theoretical models of galaxy evolution (e.g. Benson et al.\
2000; Baugh et al.\ 2005; Bower et al.\ 2006) and a direct input into
these models.

In these proceedings we outline the current status and future plans of
HiZELS. In Section 2, we describe the observational strategy. In Section
3, we show the constraints obtained on the H$\alpha$ luminosity function
and cosmic star-formation rate, and discuss the other scientific results
to date. We outline our on-going scientific work using HiZELS, and discuss
future plans, in Section 4. Section 5 presents brief conclusions.
\bigskip
\smallskip

\noindent{\large{\bf 2. Observations and sample selection}}
\medskip

\noindent{\it 2.1 Observation strategy and fields}
\smallskip

HiZELS uses observations through narrow-band filters in the $J$, $H$ and
$K$-bands (NB$_J$, NB$_H$, H$_2$(S1), with central wavelengths of 1.211,
1.619 and 2.121$\mu$m respectively), using WFCAM on UKIRT. Coupled with
broad-band filter observations, these are used to detect star-forming
galaxies in H$\alpha$ at redshifts 0.84, 1.47 and 2.23, over several
degree-scale regions of the extra-galactic sky. Of course, narrow-band
surveys are not sensitive to only one emission line, but will detect many
different emission lines at different redshifts, redshifted into the
filter. To achieve many of our goals it is necessary to identify which of
the emission-line objects are indeed H$\alpha$ emitters. HiZELS is
achieving this by targeting the best-studied regions of the extragalactic
sky, in which a wealth of multi-wavelength data already
exists. Photometric redshifts and colour-selections are being backed up by
statistical analysis of the contamination rates derived from follow-up
spectroscopy of sub-samples of emitters, as described below.

The presence of other emission line samples within the filters is actually
one of the strengths of HiZELS. The custom-made NB$_{\! H}$ and NB$_{\!
J}$ filters were specially designed such that the [O{\sc iii}]\,5007 and
[O{\sc ii}]\,3727 lines would fall into those filters for galaxies at
$z=2.23$.  This provides both a confirmation of the redshifts for a subset
of the $z=2.23$ H$\alpha$-selected sample, as well as allowing a first
investigation of the emission line properties of these sources. Other
emission lines of interest for these filters include the possibility of
detecting Ly$\alpha$ emission from galaxies at $z=8.90$ in the NB$_{\! J}$
filter.

HiZELS was awarded 22 nights on UKIRT over Semesters 07B to 09B for its
first phase (of which roughly one-third has been lost to bad weather), and
has a provisional allocation of 23 clear nights for a second phase, during
semesters 10A to 12A (subject to UKIRT remaining operational). The
original survey strategy involved observations of each field in each of
the three filters for a total on-sky observing time of $\approx
20$\,ks/pix, in order to obtain uniform coverage down to a line flux of
$10^{-16}$\,erg\,s$^{-1}$cm$^{-2}$ in each filter. In addition, a single
deeper paw-print (ie. 0.2 sq. deg., to 65ks/pix depth) using the H$_2$(S1)
filter has been taken to probe further down the H$\alpha$ luminosity
function and assess the completeness of the shallower but wider survey. As
the survey has progressed, the observing strategy has been slightly
modified based on the survey results, to an overall aim of observing as
close as possible to 1000 H$\alpha$ emitters at each redshift, split
roughly half-and-half above and below the break of the luminosity function
at each redshift. To achieve this, additional deeper paw-print
observations are proposed using the H$_2$(S1) and NB$_{\! H}$ filters,
whilst the depth of the standard survey observations using the NB$_{\! J}$
and NB$_{\!  H}$ filters, beyond the first two fields, has been
reduced. The full set of proposed observations for HiZELS, along with the
current observation status, is provided in Table~1.

\begin{center}
{\it Table 1: the HiZELS target fields, proposed exposure times, and
current observational status.~~~~~~~~~~~}
\smallskip

{\small
\begin{tabular}{lcccccccc}
  \hline
\smallskip
{\bf Field Name} & {\bf Area} & \multispan{3}{\bf Target exposure time (ks/pix)} 
            & \multispan{4}{\bf ~~Completion (\%, Dec 2009)} \\
    & {\bf [sq. deg.]} & {\bf ~~NB$_{\! J}$~~} & {\bf ~~NB$_{\! H}$~~} &
            {\bf H$_2$(S1)} && 
            {\bf ~~NB$_{\! J}$~~} & {\bf ~~NB$_{\! H}$} & {\bf H$_2$(S1)} \\
\hline
UKIDSS UDS    & 0.8 & 20.0 & 20.0 & 20.0 && 100\% & 100\% & 100\% \\
COSMOS-1$^*$  & 0.8 & 20.0 & 20.0 & 20.0 && 100\% &  50\% & 100\% \\
COSMOS-2$^*$  & 0.8 &  3.0 & 14.0 & 20.0 &&   0\% &   0\% &   0\% \\
ELAIS N1      & 0.8 &  3.0 & 14.0 & 20.0 &&   0\% &  38\% & 100\% \\
SA\,22        & 0.8 &  3.0 & 14.0 & 20.0 &&   0\% &   0\% &  50\% \\
Bo\"otes      & 0.8 &  3.0 & 14.0 & 20.0 &&   0\% &   0\% &   0\% \\
Lockman Hole  & 0.8 &  3.0 & 14.0 & 20.0 &&   0\% &   0\% &   0\% \\    
\hline				  	 		  	   
COSMOS-DeepK  & 0.4 &      &      & 65.0 &&       &       &  50\% \\ 
COSMOS-DeepH  & 0.2 &      & 50.0 &      &&       &   0\% &       \\
\hline
\end{tabular}

$^*$ -- two WFCAM pointings will be placed inside the 2 sq. deg. COSMOS
field.
}
\end{center}

\bigskip

\noindent{\it 2.2 Selection of narrow-band H$\alpha$ emitters}
\medskip

Combining the narrow-band observations with broad-band observations of the
same fields (either our own dedicated observations, or archival data from
the UKIDSS survey), narrow-band emitters are selected according to the
following criteria: (i) the object must be robustly detected on the
narrow-band image, with a signal-to-noise (S/N) above 3, in a 3-arcsec
diameter aperture; (ii) it must present a `broad-band minus narrow-band'
colour excess with a significance $\Sigma \ge 2.5$; (iii) the line
emission must have equivalent width above 50\AA; (iv) the object must be
visually confirmed as reliable, and not associated with any cross-talk
artefact. The left panel of Figure 1 shows these selection criteria for
the NB$_J$ observations of the COSMOS-1 pointing. For more details on
these selections, see Geach et~al.\ (2008; hereafter G08) or Sobral et al
(2009a; hereafter S09a). S09a have shown that these criteria are very
robust.

Our observations identify approximately 800, 350 and 300 narrow-band
emitters per pointing (0.8 sq. deg.) in the NB$_J$, NB$_H$, and H$_2$(S1)
observations, respectively. Photometric redshifts and colour selections
are then used to identify which of the emission line galaxies are
H$\alpha$.  We are able to successfully recover relatively clean H$\alpha$
samples, since strong contaminating emission lines are sufficiently well
spread in wavelength from H$\alpha$ that photometric redshifts do not have
to be very precise: $\delta z/z\sim 0.5$ -- more details can be found in
G08, S09a, Sobral et~al.\ (2010c), and Geach et~al.\ (2010). In the
NB$_{\! J}$ observations, photometric redshifts indicate that over half of
the detected emitters are indeed H$\alpha$ emitters at $z=0.84$, with a
significant fraction of the remainder being H$\beta$ or [OIII] emitters at
$z \sim 1.4$ (see middle panel of Figure 1). Archival spectroscopic
redshifts for over 100 of the emitters confirm the high completeness and
reliability of the photometric selection (Figure~1, right panel; see also
S09a). In the COSMOS and UDS fields, photometric redshifts provide a
similar level of accuracy for selecting H$\alpha$ emitters at $z=1.47$
from the NB$_H$ observations (see Sobral et~al.\ 2010c), and approximately
half of the narrow-band emitters are associated with H$\alpha$. The
H$_2$(S1) observations suffer considerably more contamination from lower
redshift emitters (e.g. Paschen and Brackett series; see G08), but are
producing around 90 candidate $z=2.23$ sources per field.

\begin{tabular}{ccc}
 \hspace*{-0.7cm}  \includegraphics[angle=0,width=57mm]{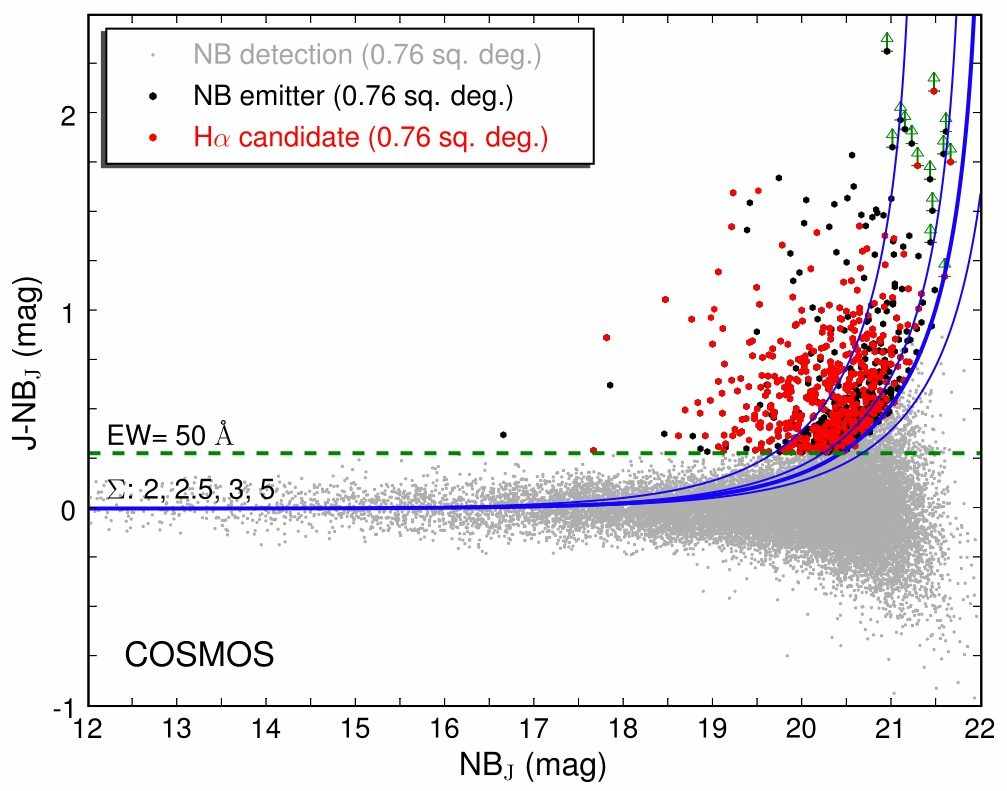} &
 \hspace*{-0.5cm}\raisebox{1mm}{\includegraphics[angle=0,width=55mm]{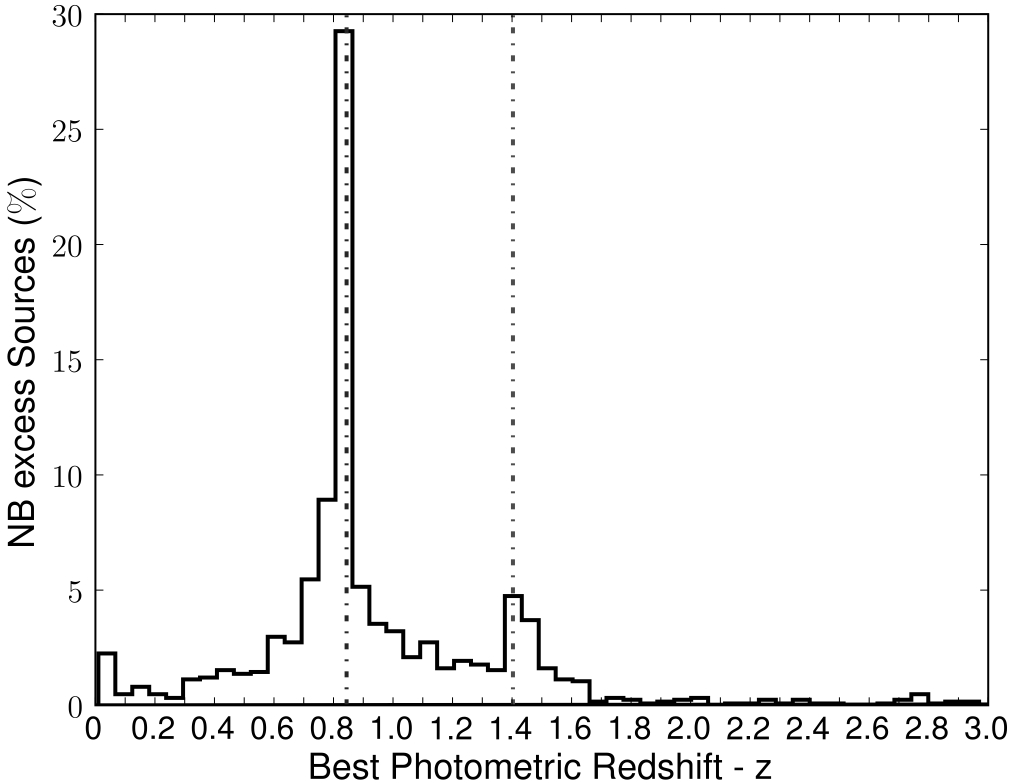}} & 
 \hspace*{-0.5cm}  \includegraphics[angle=0,width=60mm]{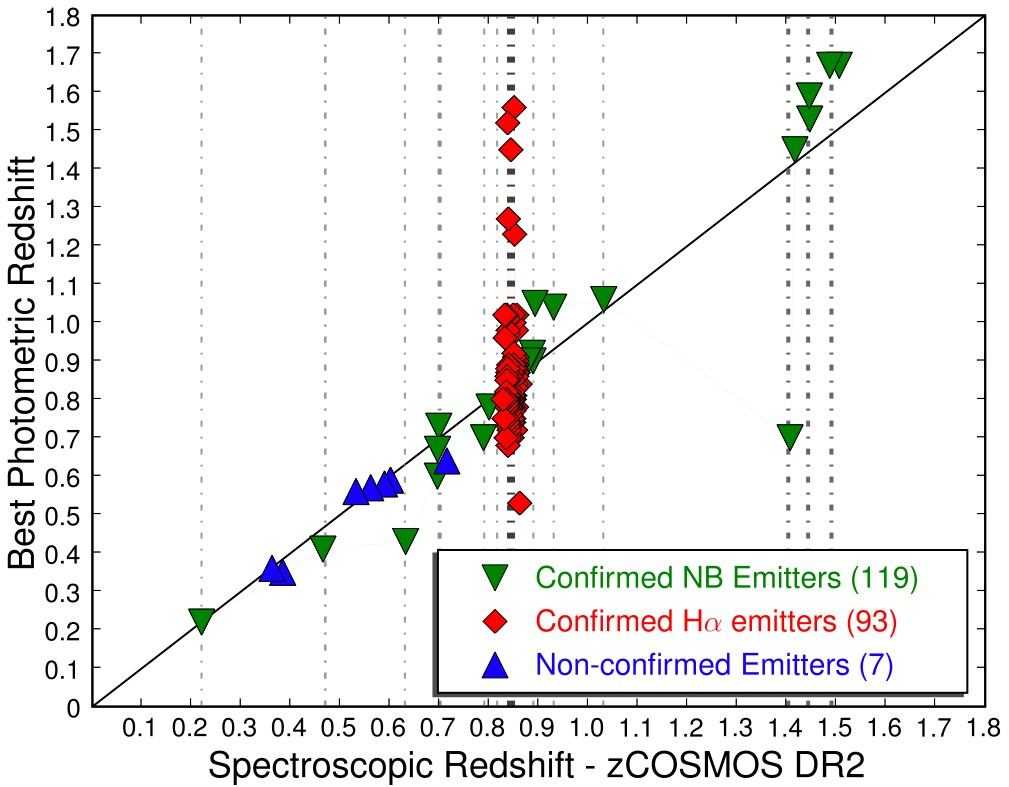} \\
\end{tabular}

%
%
\noindent\parbox{\textwidth}{\it \small Fig.~1. Left: A
colour-magnitude plot demonstrating the selection of narrow-band excess
sources (adopted from S09a).  All $>3$-$\sigma$ detections in the NB$_{\!
J}$ image are plotted and the curves represent $\Sigma$ significances of
5, 3, 2.5 and 2, respectively. The dashed line represents an equivalent
width cut of 50\AA. All selected narrow-band emitters are plotted in
black, while candidate H$\alpha$ emitters (selected using photometric
redshifts) are plotted in red. Middle: The distribution of photometric
redshifts of the NB$_{\! J}$ excess sources, showing clear peaks for
H$\alpha$ at $z=0.84$ and H$\beta$ or [OIII] at $z \sim 1.4$. Right: a
comparison between photometric and archival spectroscopic redshifts,
demonstrating the reliability of the sample.}
\bigskip
\smallskip

\noindent{\bf 3 Scientific results from HiZELS}
\medskip

\noindent{\it 3.1 The H$\mathbf{\alpha}$ luminosity function and the
  cosmic star-formation rate density}
\smallskip

HiZELS has already resulted in by far the largest and deepest survey of
emission line selected star-forming galaxies at each of the three targeted
redshifts, and has greatly improved determinations of the H$\alpha$
luminosity function. It has produced the first reliable H$\alpha$ LF at
$z=2.23$ (G08; Geach et al.\ 2010), as well as providing the first
statistically significant samples at redshifts 0.84 (S09a) and 1.47
(Sobral et al.\ 2010c). The luminosity functions are derived after
correcting the observations for: (i) contamination of the emission line
flux by the nearby [NII] line (using the relation between the flux ratio
$f_{[NII]}/f_{H\alpha}$ and the total measured equivalent width; cf S09a);
(ii) extinction of the H$\alpha$ emission line, taken to be the canonical
value of 1 magnitude (but see Section~3.5 for more details on this); (iii)
the detection completeness of faint galaxies, and the selection
completeness for detected galaxies with faint emission lines (evaluated
through Monte-Carlo simulations); (iv) filter profile effects, due to the
filter not being a perfect top-hat (again, evaluated through Monte-Carlo
simulations). For more details see G08 and S09a.

The derived luminosity functions show very strong evolution from redshift
zero out to $z=2.23$ (G08, S09a; left panel of Figure~2). At all
redshifts the H$\alpha$ LF is found to be well-fitted by a Schechter
function, but the form of the LF undergoes dramatic evolution through the
redshift range probed by HiZELS. $\phi^*$ and $L^*$ both evolve strongly
from the local Universe out to $z\sim1$, but beyond that L$^*$
continues to rise up to $z\sim2$ whilst $\phi^*$ peaks at $z\sim1$ and
then decreases at higher redshifts. The integrated luminosity function,
corrected for AGN contamination, is used to estimate the cosmic
star-formation rate density ($\rho_{SFR}$) at each redshift; using a
single star-formation tracer (H$\alpha$) from $z=0$ to $z=2.23$,
$\rho_{SFR}$ is found to rise strongly up to $z\sim1$ and then appears to
flatten out to $z\sim2.2$ (right panel of Figure~2).

%
%
\begin{center}
\begin{minipage}{145mm}
  \includegraphics[angle=0,width=145mm]{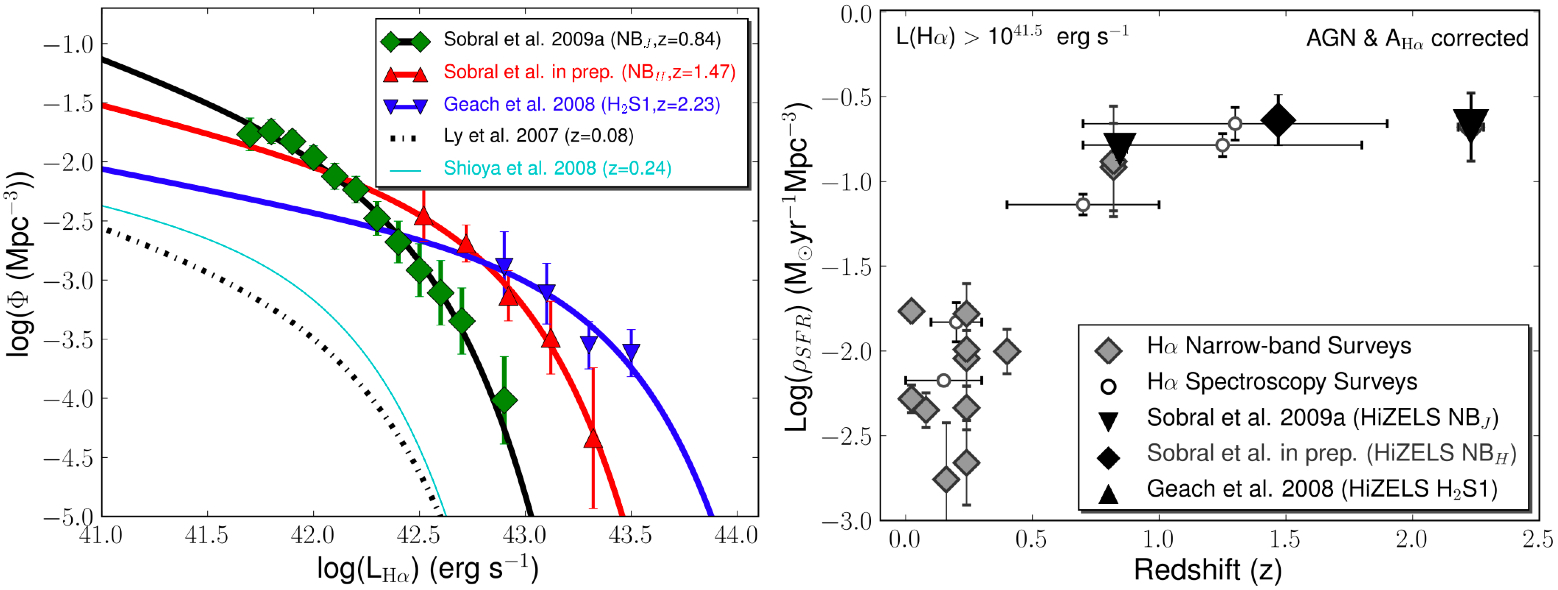}
\end{minipage}
\end{center}

\noindent\parbox{\textwidth}{ \it \small Fig.~2. Left: the $z=0.84$,
$z=1.47$ and $z=2.23$ H$\alpha$ luminosity functions from HiZELS
(corrected for [N{\sc ii}] contamination, completeness, extinction and
filter profile biases) with the best-fit Schechter functions overlaid.
Other luminosity functions from H$\alpha$ surveys at different redshifts
are presented for comparison, showing a clear evolution of the LF out to
$z > 2$.  Right: the evolution of $\rho_{SFR}$ as a function of redshift
based on H$\alpha$ (down to the HiZELS limit). This shows a rise in
$\rho_{SFR}$ up to at least $z\sim 1$, slightly steeper than the canonical
$(1+z)^4$, followed by a flattening out to at least $z\sim 2.2$.}
\bigskip

\noindent{\it 3.2 The morphologies of the H$\mathbf{\alpha}$ emitters}
\smallskip

\begin{wrapfigure}[16]{r}{0.5\textwidth} 
\vspace*{-1cm}

\begin{center}
\includegraphics[angle=0,width=75mm]{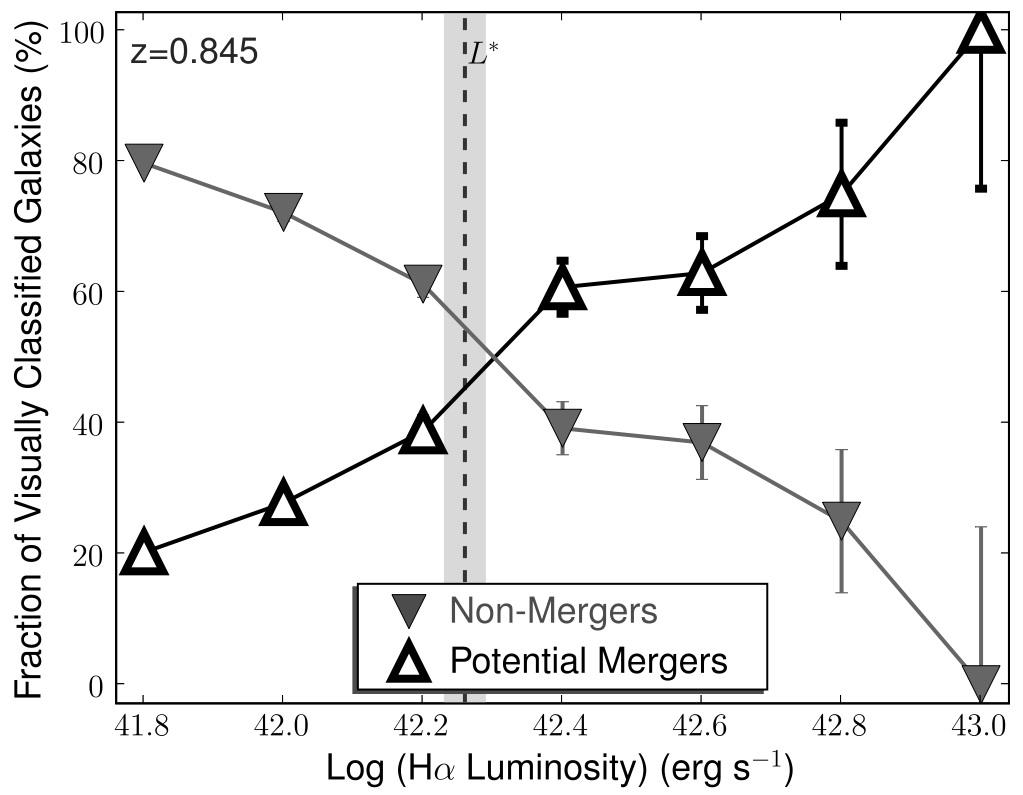}
\parbox{0.45\textwidth}{ \it\small Fig.~3. The merger
  fraction as a function of H$\alpha$ luminosity at $z=0.84$. This shows a 
clear dependency, with mergers dominating above L$^*$ and passive
  quiescent galaxies dominating below that.}
\end{center}
\end{wrapfigure}

At $z=0.84$, the H$\alpha$ emitters are mostly morphologically classed as
disks, with irregulars and mergers forming a much smaller fraction of the
sample (S09a). A strong relation is found between morphology and H$\alpha$
luminosity, however, with the fraction of irregulars/mergers rising
steadily with luminosity and the fraction of quiescent disks falling; the
break of the luminosity function seems to define a critical switch-over
luminosity between the two populations. Out to $z\sim1$, the integrated
$\rho_{SFR}$ is produced predominantly by disk galaxies and it is their
evolution which drives the strong increase in the cosmic star-formation
rate density from the current epoch to redshift one. In contrast, the
continued strong evolution of $L^*$ between $z=0.84$ and $z =2.23$
suggests an increasing importance of merger-driven star-formation activity
beyond $z\sim1$, as mergers and irregulars dominate the bright end of the
luminosity function. Analysis of the first $z=2.23$ H$\alpha$ emitters
(G08) indicates that these do show a range of morphologies, and that
indeed many show evidence of on-going merger activity. The completed
HiZELS survey will provide statistically significant samples of H$\alpha$
emitters at $z=2.23$ and $z=1.47$, allowing a direct test of whether the
change in the form of the luminosity function at $z \sim 1$ is driven by
the different evolutionary behaviour of these two different populations of
star-forming galaxies.
\bigskip

\noindent{\it 3.3 Clustering of star-forming galaxies and environmental
  variations} 
\smallskip

HiZELS is ideal for investigating the clustering of SF galaxies because
the narrow-band selection removes almost all of the projection effects
which usually degrade clustering analysis based on imaging data. The
HiZELS emitters are observed to be significantly clustered at all
redshifts.  At $z=0.84$ the characteristic correlation length calculated
from the H$\alpha$ emitters is r$_0=2.7 \pm 0.3\quad h^{-1}$\,Mpc (Sobral
et~al. 2010a), consistent with them residing in dark matter halos of mass
$\sim 10^{12}$\,M$_\odot$ at that epoch, close to that expected for the
progenitors of galaxies like the Milky Way. Using the large sample
available at this redshift, the clustering is found to be strongly
dependent on H$\alpha$ luminosity, and also to increase more weakly with
near-infrared luminosity (roughly tracing stellar mass). The clustering
amplitude is found to be independent of morphology, once the dependencies
of morphology on stellar mass and star-formation rate have been accounted
for. These results, shown in Figure 4, are qualitatively in line with
those found in the nearby Universe.

%
%
\begin{tabular}{ccc}
 \hspace*{-0.8cm} \includegraphics[angle=0,width=60mm]{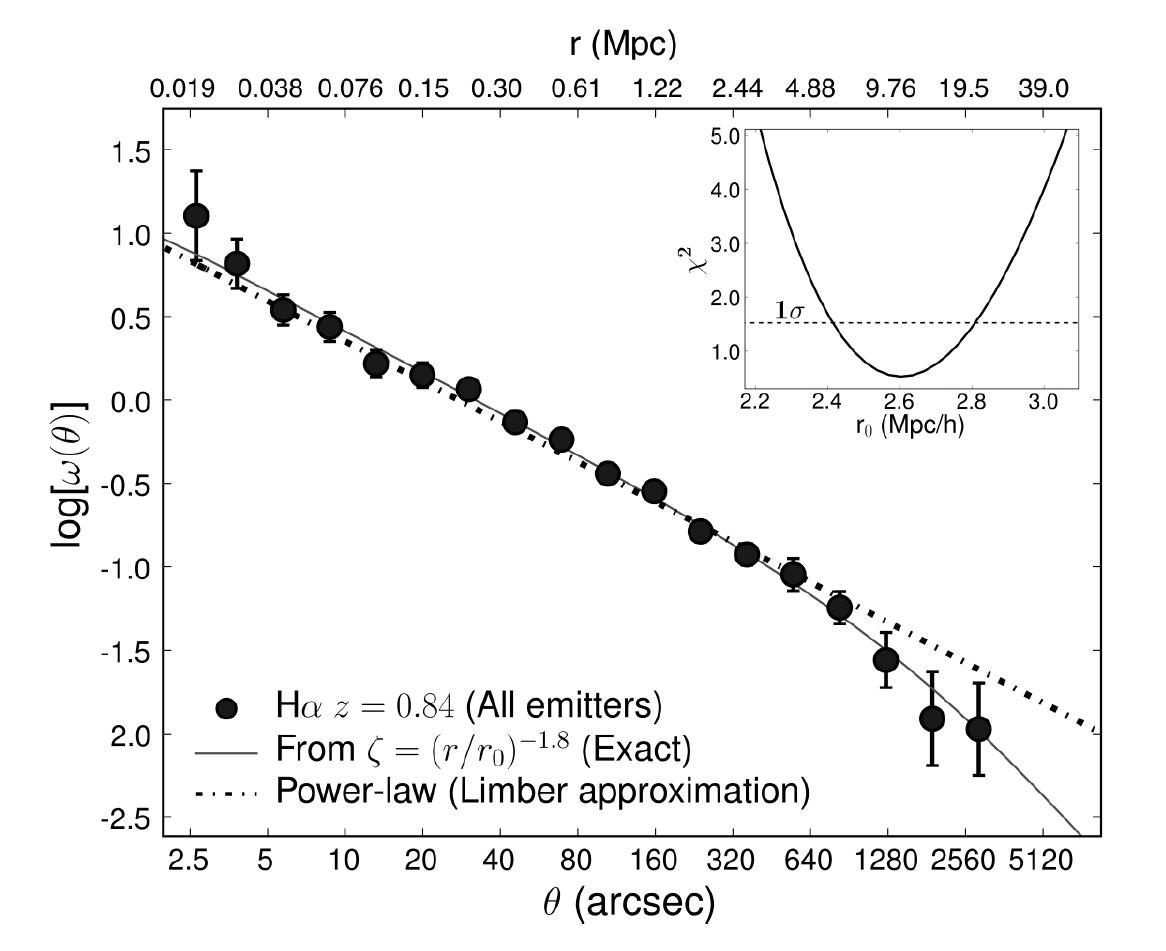}
 \hspace*{-0.5cm} \raisebox{-1mm}{\includegraphics[angle=0,width=60mm]{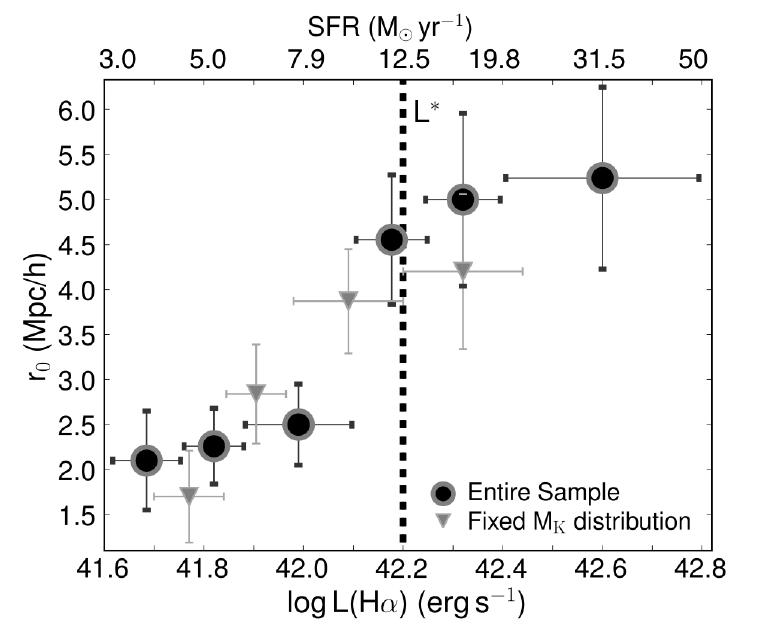}}
 \hspace*{-0.5cm} \raisebox{-2mm}{\includegraphics[angle=0,width=60mm]{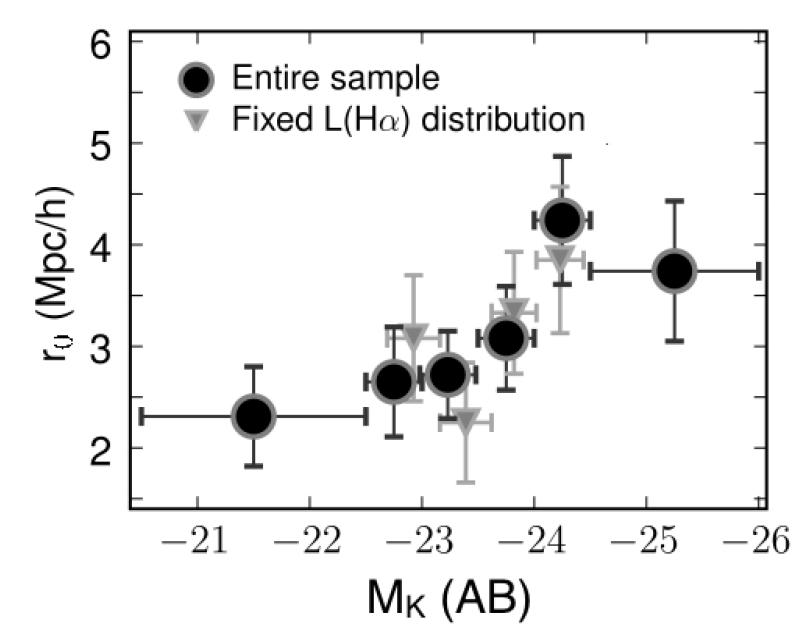}}
\end{tabular}

\noindent\parbox{\textwidth}{ \it \small Fig.~4. Left: the angular
  cross-correlation function of H$\alpha$ emitters at redshift $z=0.84$ in
  the COSMOS and UDS fields. Middle, right: the dependence of the clustering
  amplitude at this redshift on star-formation rate and absolute K-band
  magnitude.
}
\smallskip

The left-panel of Figure~5 shows the clustering amplitude of H$\alpha$
emitters measured from narrow-band surveys at different redshifts.
However, the strong dependence of the clustering amplitude on H$\alpha$
luminosity implies that considerable care must be taken when attempting to
compare these, since they have very different luminosity limits. The
right-hand panel of Figure~5 compares the clustering amplitude of the
H$\alpha$ emitters, as a function of H$\alpha$ luminosity, at three
different redshifts: $z=0.84$ and $z=2.23$ from HiZELS (Sobral
et~al. 2010a, and G08), and $z=0.24$ from Shioya et~al. (2008). It can be
seen that at a given halo mass, star-formation is much more efficient
(higher $L_{H\alpha}$) at higher redshifts.

\begin{center}
\begin{minipage}{150mm}
  \includegraphics[angle=0,width=150mm]{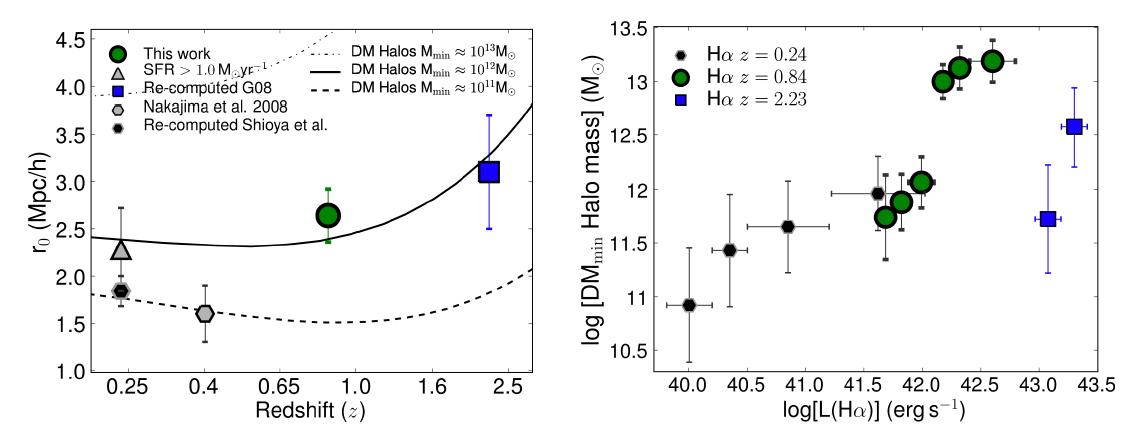}
\end{minipage}
\end{center}

\noindent\parbox{\textwidth}{ \it \small Fig.~5.  Left: The clustering
length ($r_0$) as a function of redshift for H$\alpha$ emitters selected
by narrow-band surveys. The H$\alpha$ emitters at $z=0.84$ and $z=2.23$
studied by HiZELS reside in typical dark matter haloes of $M_{\rm min}
\approx 10^{12} M_{\odot}$, consistent with being the progenitors of
Milky-Way type galaxies.  The lower luminosity H$\alpha$ emitters found in
smaller volumes at $z=0.24$ and $z=0.4$ reside in less massive
haloes. Right: The minimum mass of host dark matter haloes as a function
of H$\alpha$ luminosity at three different redshifts. From Sobral et~al.\
(2010a).}
\bigskip

The increase in $L_{H\alpha}$ at given halo mass with redshift nearly
exactly mirrors that of the increase in L$^*$ of the H$\alpha$ luminosity
function. This implies that galaxies in a given dark matter halo mass may
form stars at the same fraction of the characteristic star-formation rate
at that redshift, at all epochs (cf. Sobral et~al. 2010a). This in turn
would suggest a fundamental connection between the strong negative
evolution of the H$\alpha$ L$^*$ since $z \sim 2$ and the quenching of
star-formation in galaxies within haloes significantly more massive than
$10^{12}M_{\odot}$.
\bigskip

\noindent{\it 3.4 The masses and environments of star-forming galaxies}
\smallskip

In the nearby Universe, the stellar populations of the most massive
galaxies are observed to have formed earlier than those of less massive
galaxies. The HiZELS sample allows a direct investigation of the redshift
at which this ``down-sizing'' process began, and a measurement of how the
characteristic stellar mass of star-forming galaxies varies with
redshift. At $z=0.84$ we find a strong dependence of the proportion of
galaxies which are actively forming stars as a function of mass: for $M
\approx 10^{10}\,M_\odot$, around a third of galaxies are detected by
HiZELS, but at higher masses ($>10^{11.5}$\,M$_\odot$) this fraction falls
away to a value consistent with zero (Sobral et~al. 2010b). This indicates
that down-sizing is already in place at $z=0.84$.

\begin{wrapfigure}[18]{r}{0.45\textwidth} 
\vspace*{-1cm}

\begin{center}
\includegraphics[angle=0,width=70mm]{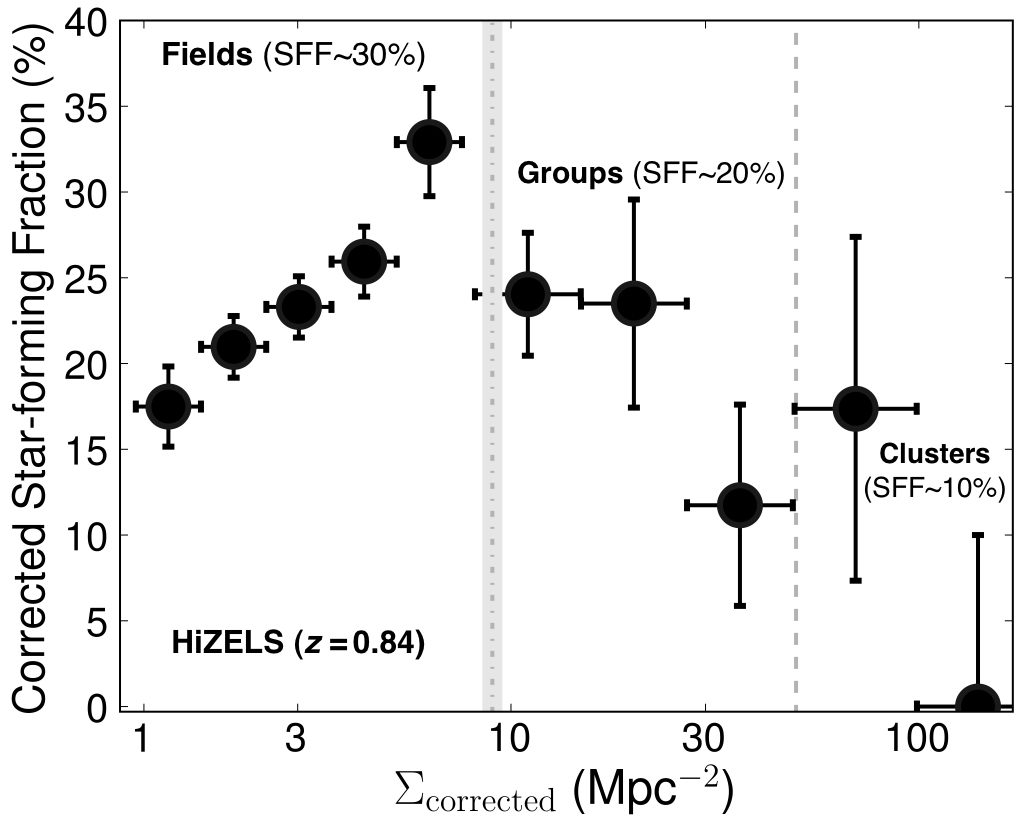}
\parbox{0.45\textwidth}{ \it\small Fig.~6. The fraction of galaxies
forming stars as a function of local galaxy surface density, for
H$\alpha$ emitters at $z=0.84$. The star-forming fraction increases
with local surface density in the field, but then decreases in 
group and cluster environments.}
\end{center}
\end{wrapfigure}

We find that at $z=0.84$ the fraction of star-forming galaxies increases
with local galaxy surface density at low galaxy surface densities, but
then falls in group and cluster environments (Figure~6). The median star
formation rate of the H$\alpha$ emitters increases with local galaxy
density: those residing in denser regions are mostly starbursts while the
H$\alpha$ emitters found in less dense regions present much more quiescent
star-formation. Our results resolve the apparent contradictions between
different studies presented in the literature, which have found the
star-forming galaxy fraction to increase (e.g. Elbaz et al 2007, studying
field galaxies) or decrease (e.g. Patel et al 2009, studying galaxies in
the high density regions around a rich cluster) with increasing
environmental density: both trends exist, and which is observed depends
upon the range of environmental densities studied.

With the full HiZELS sample we will be able to investigate how these
relations with mass and environment evolve to higher redshifts, over 80\%
of the lifetime of the Universe
\medskip

\noindent{\it 3.5 The dust extinction properties of high-z star-forming 
galaxies}
\smallskip

Deep 24-$\mu$m data from {\it Spitzer} are available in most of the HiZELS
fields, and 35\% of the $z=0.84$ HiZELS H$\alpha$ emitters in UDS and
COSMOS are detected at $24\mu$m. Using these detections and stacking
analyses, the star-formation rate estimates from the HiZELS H$\alpha$
sample can be compared with those from the mid-infrared to estimate the
dust extinction properties of the star-forming galaxies. We find a clear
trend for an increase in mean dust extinction with increasing star
formation rate at $z=0.84$ (Garn et al.\ 2009; see Figure~7). The relation
we determine broadly matches that found in the low redshift Universe by
Hopkins et~al.\ (2001), suggesting that there is no significant change in
the dust properties of star-forming galaxies with redshift, at least out
to $z \sim 1$.  We find no variation of the extinction--SFR relation with
galaxy morphology, environment or merger status. Carrying out equivalent
analyses at the higher HiZELS redshifts will be a key goal when the
samples are sufficiently large.

We are exploiting the exquisite multi-wavelength data available for our
fields to understand the spectral energy distribution of typical
star-forming galaxies across a range of redshifts. We aim to compare
additional star-formation tracers (UV continuum, [OII], sub-mm, radio)
with our H$\alpha$ measurements.
\medskip

\begin{center}
\begin{minipage}{150mm}
  \includegraphics[angle=0,width=150mm]{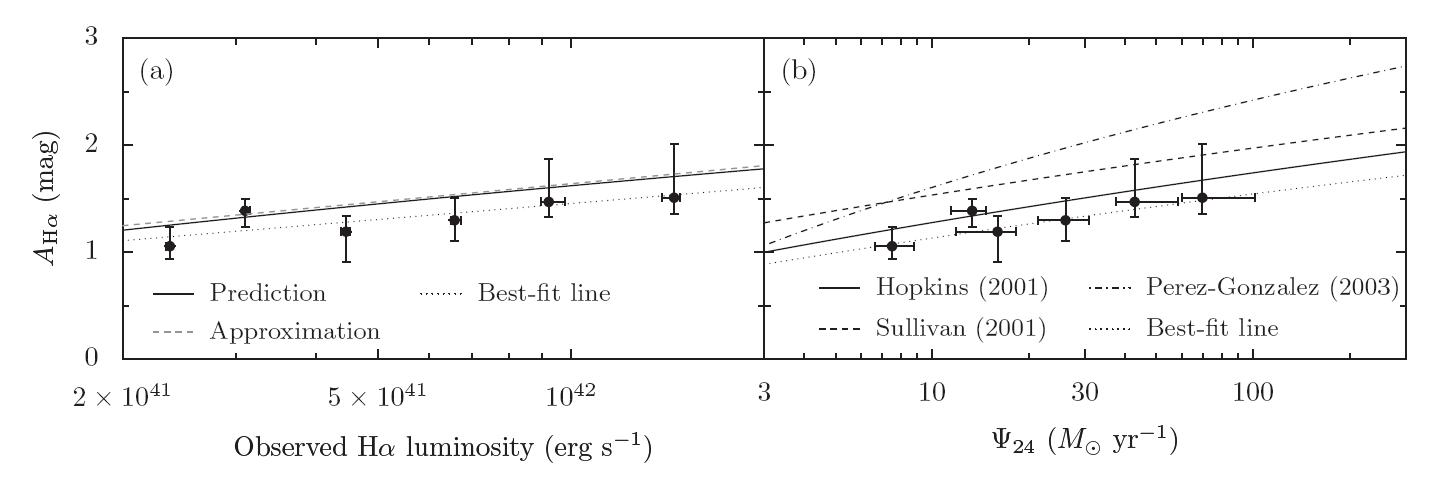}
\end{minipage}
\end{center}

\noindent\parbox{\textwidth}{ \it \small Fig.~7. The variation in
H$\alpha$ extinction for H$\alpha$ emitters binned by their observed
H$\alpha$ flux, as a function of H$\alpha$ luminosity (left) or 24$\mu$m
star-formation rate (right). The best-fit (dotted line) shows that, at the
4$\sigma$ confidence level, extinction increases with star-formation
rate. The slope and normalisation of the relationship are comparable to
those derived in the local Universe using the fit between the Balmer
decrement and SFR, by Hopkins et al. (2001).}
\smallskip

\medskip

\noindent{\it 3.6 Limits on the space density of bright $z=8.9$ Ly$\alpha$
emitters}
\smallskip

Parallel to our H$\alpha$ survey using the NB$_{\! J}$ filter, we have
explored its capabilities to detect bright Ly$\alpha$ emitters at $z=8.9$,
when the Universe was only $\sim 0.5$\,Gyr old. Detection of any such
objects would have important consequences for our understanding of the
early star-formation history of the Universe, as well as providing an
important probe of the re-ionisation of the Universe. After conducting an
exhaustive search for such emitters in the UDS and COSMOS fields, two
candidates were isolated, both in the COSMOS field. Follow-up spectroscopy
using CGS4 in January 2009, together with follow-up $J$-band imaging
obtained in February 2009 with WFCAM has shown that these cannot be
Ly$\alpha$ emitters. These results have allowed us to improve constraints 
on the bright end of the $z \sim 9$ Ly$\alpha$ luminosity function by
three orders of magnitude (Sobral et al.\ 2009b).
\bigskip
\smallskip

\noindent{\bf 4. On-going work and future plans}
\medskip

HiZELS offers a powerful resource for a large number of additional
studies. Other on-going work includes the following:

\begin{bul2}
\item {\it Spectroscopic confirmation of emission lines and study of
emission line ratios}. We have followed-up HiZELS H$_2$(S1) H$\alpha$
emitters using near-infrared spectroscopic observations, both with
VLT/ISAAC and with Gemini. These observations have successfully confirmed
the HiZELS emission line detections, and the results will be published in
Geach et al.\ (2010). We are carrying out optical spectroscopic
observations of a large fraction of the emitters selected from all 3
narrow bands (NB$_{\! J}$, NB$_{\! H}$ \& H$_2$(S1)), in the UDS field,
using VIMOS. The main aims are to confirm redshifts, to investigate
line-ratios and extinction, to identify AGN, and to test the robustness of
the selection criteria used for HiZELS.  By allowing pre-selection of
galaxies with known emission lines in the $J$- and $H$-bands, HiZELS will
also provide a valuable input sample for early commissioning tests for
FMOS, the new near-IR multi-object spectroscope on Subaru; such
observations will allow us to confirm emission lines, redshifts, identify
AGN, and study sample contaminants.

\item {\it Complementary narrow-band observations}. Using HAWK-I on VLT we
have carried out ultra-deep H$_2$(S1) exposures in two small regions in
the COSMOS and UDS fields. These will reach a depth of $\approx 3
M_{\odot}$yr$^{-1}$ at $z=2.23$, comparable to the HiZELS NB$_{\! J}$
sensitivity, and will thus provide an excellent complement to our
shallower but much wider HiZELS imaging. This will enable an accurate
measurement of the very faint end of the H$\alpha$ luminosity function at
$z\sim2.23$, determining the global star-formation history and testing for
differential evolution of the most/least active galaxies.

\item {\it Ly$\alpha$ and H$\alpha$ emission from galaxies at $z=2.23$.}
We are exploiting existing narrow-band Ly$\alpha$ imaging of the COSMOS
field from Nilsson et al.\ (2009) to compare the Ly$\alpha$ and H$\alpha$
properties of galaxies at $z=2.23$.  Ly$\alpha$ emission has been
extensively used to identify and study large samples of star-forming
galaxies at the highest redshifts, but its reliability as a tracer of
complete galaxy populations remains untested at cosmological distances. By
combining our H$\alpha$ sample with Ly$\alpha$ observations we can derive
the ratio of Ly$\alpha$ to H$\alpha$ emission and hence obtain a direct
measurement of the escape fraction of Ly$\alpha$ photons with few model
assumptions -- something which has never been done at high redshift.  We
find that only a modest fraction, $10\pm 3$\%, of H$\alpha$ emitters are
detected in Ly$\alpha$ above a rest-frame equivalent width of 20\AA\
(Matsuda et al.\ 2010, in prep).  This detection rate is similar to that
seen for $z\sim 2.3$ UV-continuum selected galaxies (Reddy et al.\ 2009)
and is less than that seen for $z\sim 3$ Lyman-break galaxies (Shapley et
al.\ 2004).  Stacking the H$\alpha$ sources we find potential evidence for
extended Ly$\alpha$ halos, similar to that seen around $z\sim 3$ LBGs by
Hayashino et al.\ (2004).

\item {\it Paschen-series luminosity functions.} In our H$_2$(S1) imaging
we are also detecting large samples of Pa$\alpha$, Pa$\beta$ and other
lines from lower redshift sources. Whilst our photometric selection is
very efficient at filtering out these contaminants from our H$\alpha$
studies, we aim to present a robust analysis of the extent and character
of the contamination. This may allow the derivation of Pa-series galaxy
luminosity functions, which would provide a new measure of the cosmic
star-formation rate density, and an estimate of extinction.

\item {\it Kinematics of HiZELS galaxies}.  We are using the SINFONI
near-IR integral field spectrograph on VLT to map the kinematics of the
H$\alpha$ emission in a sample of 18 H$\alpha$ emitters at $z=0.84$,
$z=1.47$ and $z=2.23$ in the COSMOS field.  This project exploits HiZELS
panoramic coverage to select sources at each redshift which are matched on
H$\alpha$-luminosity and, critically, are close to natural AO guide
stars. These observations will allow us to map the distribution of
H$\alpha$ emission on $\sim 0.2''$ scales within these galaxies to search
for kinematic evidence of rotating discs, major mergers, etc. This
detailed follow-up will be a valuable addition to HiZELS, allowing us to
study the properties of the progenitors of Milky Way-type galaxies seen at
the time when their star-formation was at its peak.

\item {\it Comparison with galaxy formation models.} HiZELS provides a
large sample of star-forming galaxies at three different redshifts, the
selection of which can be cleanly replicated in (and compared with)
theoretical semi-analytic models of galaxy formation. An initial
comparison of the luminosity function and clustering properties of the
H$\alpha$ emitters at $z=2.23$ was presented in G08, showing that these
were broadly in line with those predicted by one particular ``recipe'' of
galaxy formation, that of Bower et al.\ (2006). Now with larger datasets
available at all three wavelengths, far more detailed investigations are
beginning, folding in the variations of the luminosity function and
clustering of the HiZELS galaxies with morphology, mass, etc. These
studies will offer an invaluable ingredient to these models.
\end{bul2}
\bigskip

\noindent{\bf 5. Conclusions}
\medskip

The advent of wide-field imaging cameras in the near-IR, especially WFCAM
on UKIRT, has revolutionised the study of high-redshift emission line
galaxies. HiZELS aims to detect and study up to a thousand H$\alpha$
emitting galaxies at each of three redshifts, 0.84, 1.47 and 2.23,
spanning the peak epoch of star-formation in the Universe. This goal is
already nearly completed at $z=0.84$, and has resulted in a plethora of
results related to the luminosity function, masses, morphologies,
environments, clustering, dust extinction, and other properties of the
star-forming galaxies at this redshift. Results at the higher redshifts
remain sparser to date, because the current samples are smaller due to the
lower sky density of H$\alpha$ emitters at these higher redshifts: as the
survey progresses, we will replicate our $z=0.84$ studies at the higher
redshifts, where arguably our H$\alpha$ studies will be most unique, and
the impact of HiZELS will be highest.
\bigskip

\noindent{\bf Acknowledgements}
\medskip

\noindent HiZELS is based on observations obtained with the Wide Field
CAMera (WFCAM) on the United Kingdom Infrared Telescope (UKIRT). We are
indebted to Andy Adamson, Luca Rizzi, Chris Davis, Tim, Thor and Jack for
their support at the telescope. PNB is grateful for support from the
Leverhulme Trust, DS for support from the FCT, and JEG for support from
STFC.
\medskip

This article is dedicated to the memory of Timothy Garn, a valued member
of the HiZELS team, who nature has cruelly taken away at a time when the
prime of his research career was still in front of him.
\bigskip

\noindent{\bf References}
\medskip

\noindent Baugh C.M. et al.\ 2005, MNRAS, 356, 1191 \\ 
Benson A.J., Cole S., Frenk C.S., Baugh C.M., Lacey C.G., 2000, MNRAS, 311, 793   \\ 
Best P.N., 2004, MNRAS, 351, 70\\
Best P.N., Kaiser C.R., Heckman T.M., Kauffmann G., 2006, MNRAS, 368, L67\\
Bower R. et al.\ 2006, MNRAS, 370, 645    \\ 
Cowie L.L., Songaila A., Hu E.M., Cohen J.G., 1996, AJ, 112, 839\\ 
Doherty M., Bunker A., Sharp R., Dalton G., Parry I., Lewis I., 2006,
MNRAS, 370, 331\\
Elbaz D. et~al., 2007, A\&A, 468, 33\\
Gallego J., Zamorano J., Aragon-Salamanca A., Rego M., 1995, ApJL, 455, L1\\
Garn T. et al.\ 2009, MNRAS, in press, arXiv/0911.2511 \\ 
Geach J.E., Smail I., Best P.N., Kurk J., Casali M., Ivison R.J., Coppin K., 2008, MNRAS, 388, 1473\\
Geach J. et al.\ 2010, MNRAS, in prep. \\ 
Hayashino T. et al. 2004, AJ, 128, 2073 \\ 
Hopkins A. M., Beacom J. F., 2006, ApJ, 651, 142\\
Ivison R. et al.\ 2007, MNRAS, 380, 199 \\ 
Kennicutt R.C. Jr, 1998, ARAA, 36, 189 \\ 
Kodama T. et al.\ 2007, MNRAS, 377, 1717 \\ 
Lewis I. et al.\ 2002, MNRAS, 334, 673 \\ 
Lilly S.J., Le Fevre O., Hammer F., Crampton D., 1996, ApJL, 460, L1+\\
Moorwood A.F.M., van der Werf P.P., Cuby J.-G., Oliva E., 2000, A\&A, 326, 9\\
Moustakas J., Kennicutt R.C. Jr., Tremonti C.A., 2006, ApJ, 642, 775 \\ 
Nilsson K.K. et al.\ 2009, A\&A, 498, 13 \\ 
Patel S.G., Holden B.P., Kelson D.D., Illingworth G.D., Franx M., 2009, ApJ, 705, L67\\
Reddy N.A., Steidel C.C., 2009, ApJ, 692, 778 \\ 
Shapley A.E., Erb D.K., Pettini M., Steidel C.C., Adelberger K.L., 2004, ApJ, 612, 108 \\ 
Shioya Y. et al. 2008, ApJS, 175, 128\\
Sobral D. et~al., 2009a, MNRAS, 398, 75 \\ 
Sobral D. et~al., 2009b, MNRAS, 398, L68 \\
Sobral D., Best P.N., Geach J.E., Smail I., Cirasuolo M., Garn T., Dalton
G.B., Kurk J., 2010a, MNRAS, in press, arXiv/0912.3888 \\
Sobral D., Best P., Geach J., Smail I., Cirasuolo M., Garn T., Kurk J., Dalton
G., 2010b, MNRAS, submitted \\ 
Sobral D. et al.\ 2010c, MNRAS, in prep \\ 
Tresse L., Maddox S.J., Le Fevre O., Cuby J.-G., 2002, MNRAS, 337, 369\\
Yan L., McCarthy P.J., Freudling W., Teplitz H.I., Malumuth E.M., Weymann R.J., Malkan M.A., 1999, ApJL, 519, L47\\

\end{document}